\documentclass[twoside]{article}
\usepackage{amsmath}
\usepackage{amsfonts}
\usepackage{amssymb}
\usepackage{amscd}
\usepackage{color}

\begin{document}
\thispagestyle{plain}

\begin{center}
{\Large \bf \strut
Wavelets in field theory
\strut}\\
\vspace{10mm}
{\large \bf 
Wayne Polyzou$^{abc}$
and Fatih Bulut$^{d}$
}
\end{center}
\noindent{
\small $^a$\it The University of Iowa, Department of Physics and Astronomy,
Iowa City, IA }\\ 
\noindent{
\small $^b$\it This work supported by the US Department of Energy under 
contract No. DE-FG02-86ER40286}\\
\noindent{
\small $^b$\it To be published in Few-Body Systems}\\
\noindent{
\small $^d$\it In\"on\"u University,  Department of Physics,
Malatya, Turkey}   

\markboth{
W. N. Polyzou and Fatih Bulut }
{Wavelets in field theory} 

\begin{abstract}
We discuss the use of Daubechies wavelets in discretizing 
quantum field theories.
\\[\baselineskip] 
\end{abstract}

\section{Introduction}
\label{intro}

Daubechies wavelets \cite{ref1}\cite{ref2} and 
the associated scaling functions are a basis
for the square integrable functions on the real line that have
properties that make them useful for discretizing quantum
field theories \cite{ref3}.  These properties include:
\begin{itemize}
\item [1.] The basis functions have compact support.
\item [2.] Subsets of the basis functions form locally finite 
partitions of unity.
\item [3.] The basis functions are related to fixed points of 
a renormalization group equation.
\item [4.] The basis functions are natural for treating problems 
with multiple scales. 
\item [5.] The basis functions have a limited amount of smoothness.
\end{itemize} 

Wavelets methods have been discussed for applications to quantum field
theory both from a computational \cite{ref4}
\cite{ref5}\cite{ref6}\cite{ref7}\cite{ref8} and theoretical
\cite{ref9}\cite{ref10}\cite{ref11}\cite{ref12}\cite{ref13}
perspective.  In this work we provide a general discussion of the
properties of fields smeared with compactly supported wavelets.  The
multiscale structure of the wavelet basis leads to a natural framework
for eliminating short-distance degrees of freedom, resulting in 
effective models that are amenable to computation.

\section{Scaling functions and renormalization group equations}
\label{sf}

The basis functions are constructed from a single function
using a unitary dyadic scale transformation $D$ and a unit translation operator
$T$ defined by:
\begin{equation}
\underbrace{(D f)(x) = \sqrt{2} f(2x)}_{\mbox{scale change}}  
\qquad
\underbrace{(Tf)(x) = f(x-1)}_{\mbox{translation}} .
\label{a:1}
\end{equation} 
Application of the operator $D$ to a function reduces the volume of 
its support by a factor of two and adjusts the scale to preserve
the $L^2(\mathbb{R})$ norm.  The operator $T$ is a
discrete translation.

All of the basis functions are finite linear combinations of translated
and rescaled copies of a single function, called the scaling function,
$s(x)$.  The scaling function is a solution of the scaling equation 
\begin{equation}
s (x) = D 
(\sum_{l=0}^{2K-1} h_l T^l s (x)) . 
\label{a:2}
\end{equation}
Since (\ref{a:2}) is a homogeneous equation, 
the solution is fixed by the scale fixing condition
\begin{equation}
\int s (x) dx =1. 
\label{a:3}
\end{equation}

The renormalization group structure of the scaling equation is illustrated by 
\begin{equation}
f_n (x) = \underbrace{D \underbrace{
(\sum_{l=0}^{2K-1} h_l T^l f_{n-1}(x))}_{\mathbf{block\, average}}}_{\mathbf{rescale}}.  
\label{a:4}
\end{equation} 
Comparing (\ref{a:2}) to (\ref{a:4}) shows that $s(x)$ is a fixed point 
of this renormalization group equation.
The coefficients $h_l$ in (\ref{a:2}) and (\ref{a:4}) are fixed constants 
that determine properties of the fixed point, $s(x)$.  
They are solutions of the 
following set of algebraic equations:
\begin{equation}
\sum_{n=0}^{2K-1} h_n = \sqrt{2} 
\qquad
\sum_{n=0}^{2K-1} h_n h_{n-2m} = \delta_{m0} 
\qquad
\sum_{n=0}^{2K-1} n^m (-l)^n h_{2K-1-n} = 0   \qquad m < K 
\label{a:7}
\end{equation} 
where $K$ is an integer that defines the type of Daubechies scaling function.
The first equation is a necessary condition 
for the scaling equation to have a solution, 
the second equations ensure that unit translates of the scaling function are 
orthonormal, and the third equations ensure that $x^m$ for $m<K$ can be 
pointwise represented by a locally finite linear combination of  
scaling functions.
%
%
%
Exact solutions of equations (\ref{a.7}) for $K=1,2$ and $3$ can be found
in \cite{ref3}. 
The choice of $K$ controls the support, smoothness, and the number of non-zero
translated scaling functions at any point.

Scaling functions with different resolutions are obtained by applying 
integer translations and dyadic rescaling to the original scaling function
\begin{equation}
s^k_n (x) := (D^k T^n s)(x) = \sqrt{2^k} s( 2^k( x -n/2^k) ).  
\label{a:9}
\end{equation}

It can be shown that the scaling functions have the following properties:
\begin{itemize}
\item [1.] Reality: \qquad
$s^k_n (x) = s^{k}_n(x)^*$
\item [2.] Compact support: \qquad
$\mbox{support}[s^k_n (x)] = 
1/2^k [n,2K-1+n] $
\item [3.] Orthonormality: \qquad
$\int s^k_m(x)s^k_n(x) dx   = \delta_{mn}$
\item [4.] Pointwise low-degree polynomial representation:\qquad 
$x^m = \sum_n c_n(m) s^k_n (x) \qquad m < K,\, \mbox{any} \, k$
\item [5.] Partition of unity: \qquad
$1 = \sum_{n=-\infty}^{\infty} \sqrt{{1 \over 2^k}} s^k_n (x-n) \qquad \mbox{any} \, k  $ 
\item [6.] Differentiability (K$>$2): \qquad
${ds^k_n (x ) \over d x } \qquad \mbox{exists}
\qquad \mbox{for} \qquad  K \geq 3$ .
\end{itemize}

The scale-$1/2^k$ scaling functions, $\{ s^k_n (x)\}_{n=-\infty}^\infty$, 
are an orthonormal basis for the resolution 
$1/2^k$ subspace, ${\cal S}_k$, of $L^2(\mathbb{R})$ defined by
\begin{equation}
{\cal S}_k := \{f(x) \vert 
f(x) = \sum_{n=-\infty}^{\infty} c_n s^k_n (x), 
\quad
\sum_{n=-\infty}^{\infty} \vert c_n \vert^2 < \infty \} .
\label{a:16}
\end{equation}

\section{Wavelets and the multiresolution decomposition of the Hilbert space
}
\label{wav}

The scaling equation implies that the $s^k_n (x)$ are
linear combinations of $s^{k+1}_n (x)$ leading to 
the nested inclusions
\begin{equation}
L^2 (\mathbb{R}) \supset \cdots 
\supset {\cal S}_{k+1} \supset  {\cal S}_{k} \supset  {\cal S}_{k-1} 
\supset \cdots .
\label{a:17}
\end{equation}

The wavelet subspace, ${\cal W}_n$, is defined as the orthogonal 
complement of ${\cal S}_n$ in ${\cal S}_{n+1}$:
\begin{equation}
{\cal S}_{n+1} = {\cal S}_{n} \oplus {\cal W}_{n}. 
\label{a:18}
\end{equation}
Repeated application of (\ref{a:18}) gives a decomposition of
${\cal S}_n$ into orthogonal subspaces with resolutions
$1/2^{n-m} \cdots 1/2^n$: 
\begin{equation}
{\cal S}_n = {\cal W}_{n-1} \oplus {\cal W}_{n-2} 
\oplus \cdots \oplus
{\cal W}_{n-m} \oplus {\cal S}_{n-m}.  
\label{a:19}
\end{equation}
This can be extended to all scales using
\begin{equation}
L^2(\mathbb{R})= \lim_{n \to \infty} {\cal S}_n = 
{\cal S}_k \oplus {\cal W}_k \oplus {\cal W}_{k+1} \oplus \cdots 
\label{a:20}
\end{equation}
which gives a decomposition of $L^2 (\mathbb{R})$
into a direct sum of orthogonal 
subspaces of different resolutions.
Orthonrmal bases for the wavelet spaces are constructed 
from the {\it mother wavelet}, defined by
\begin{equation}
w(x) = 
D \sum_{l=0}^{2K-1}g_l  T^l s(x)  \qquad 
\mbox{where}
\qquad
g_l:=(-)^l h_{2K-1-l}.
\label{a:21}
\end{equation}
The scale $k$ wavelets, 
$w^k_n (x) := D^k T^n w (x)$, are 
an orthonormal basis for the subspace ${\cal W}_k$:
\begin{equation}
{\cal W}_k := \{f(x) \vert 
f(x) = \sum_{n=-\infty}^{\infty} c_n w^k_n (x), 
\quad
\sum_{n=-\infty}^{\infty} \vert c_n \vert^2 < \infty \}.
\label{a:23}
\end{equation}

While the scaling functions can be used to point-wise represent
low-degree polynomials, the wavelets are orthogonal to the same
low-degree polynomials.
\begin{equation}
\int w^k_n(x) x^m dx = 0 \qquad 0 < m < K .
\label{a:24}
\end{equation}

Equation (\ref{a:20}) means that the scale $1/2^k$ scaling 
functions and the 
wavelets on all finer scales,
\begin{equation}
\{ s^k_n(x) \}_{n=-\infty}^{\infty} \cup
\{ w^{k+l}_n (x) \}_{n=-\infty}^{\infty}{}_{l=0}^{ \infty}
\label{a:25}
\end{equation}
are an orthonormal basis for the space of square integrable functions
on the line.  This basis has all of the properties listed in the 
beginning of these proceedings.

The relation (\ref{a:18})  
means that $\{ s^k_n (x) \}_{n=-\infty}^\infty$ and
$\{ s^{k-1}_n (x) \}_{n=-\infty}^\infty \cup
\{ w^{k-1}_n (x) \}_{n=-\infty}^\infty$ are two different orthonormal 
bases on the same space. They are necessarily related by a real orthogonal
transformation called the wavelet transform that is given explicitly by: 
\begin{equation}
s^{k-1}_n (x)   = \sum_{l=0}^{2K-1} h_l  s^k_{2n+l} (x) \qquad
w^{k-1}_n (x)   =  \sum_{l=0}^{2K-1} g_l s^k_{2n+l} (x)
\label{a:27}
\end{equation}
\begin{equation}
s^k_n(x)= \sum_{m} h_{n-2m} s^{k-1}_m (x) + 
\sum_m  g_{n-2m} w_m^{k-1}(x).
\label{a:28}
\end{equation}
Because the transformation is orthogonal, the same coefficients 
appear in the transformation and its inverse.

These relations express a fine-scale basis in terms of 
a coarse-scale basis, which is related to the fine-scale basis by 
a unitary scale transformation, and additional functions (wavelets) 
that fill in the missing 
fine-scale information.  These concepts can be used decompose local fields 
into linear combinations of discrete fields of different resolutions.

For the purpose of illustration we consider fields in $1+1$ dimensions.  
Fields on multidimensional spaces can be treated using products of 
one-dimensional scaling functions and wavelets. 
We expand a set of local fields 
$\pmb{\Phi} (x,t)$, and $\pmb{\Pi}(x,t)$ 
satisfying canonical equal time commutation relations:
\begin{equation}
[\pmb{\Phi} (x,t),\pmb{\Pi}(y,t)] =i \delta (x-y)
\label{a:29}
\end{equation}
in the basis (\ref{a:25}).

We define discrete fields by smearing these fields with 
the basis functions (\ref{a:25})
\begin{equation}
\pmb{\Phi}_s^k (n,t) := \int s^k_n (x) \pmb{\Phi} (x,t) dx \qquad
\pmb{\Phi}_w^{l} (n,t) := \int w^l_n (x) \pmb{\Phi} (x,t) dx 
\label{a:31}
\end{equation} 
\begin{equation}
\pmb{\Pi}_s^k (n,t) := \int s^k_n (x) \pmb{\Pi} (x,t) dx \qquad
\pmb{\Pi}_w^{l} (n,t) := \int w^l_n (x) \pmb{\Pi} (x,t) dx . 
\label{a:33}
\end{equation}
The discrete fields are associated with degrees of freedom of the 
theory corresponding to different compact regions. 

\section{Multiresolution fields}
\label{mf}

The orthonormality of the basis (\ref{a:25}) means that the discrete fields 
also satisfy the equal-time commutation relations
\begin{equation}
[\pmb{\Phi}_s^k (m,t) , \pmb{\Phi}_s^k (n,t )] =0  
\qquad
[\pmb{\Pi}_s^k (m,t) , \pmb{\Pi}_s^k (n,t )] =0  
\qquad
[\pmb{\Phi}_s^k (m,t) , \pmb{\Pi}_s^k (n,t )] =i \delta_{mn}  
\label{a:35}
\end{equation}
\begin{equation}
[\pmb{\Phi}_w^{k} (m,t) , \pmb{\Phi}_w^{l} (n,t )] =0  
\qquad
[\pmb{\Pi}_w^{k} (m,t) , \pmb{\Pi}_w^{l} (n,t )] =0  
\qquad
[\pmb{\Phi}_w^{k} (m,t) , \pmb{\Pi}_w^{l} (n,t )] 
=i \delta_{mn}\delta_{kl}  
\label{a:37}
\end{equation}
\begin{equation}
[\pmb{\Phi}_w^{k} (m,t) , \pmb{\Phi}_s^{k} (n,t )] =0  
\qquad
[\pmb{\Pi}_w^{k} (m,t) , \pmb{\Pi}_s^{k} (n,t )] =0  
\qquad
[\pmb{\Phi}_w^{k} (m,t) , \pmb{\Pi}_s^{k} (n,t )] =0 . 
\label{a:39}
\end{equation}
These discrete fields form a local algebra in the sense that 
there are fields with support in arbitrarily small spatial volumes.

The exact fields can be expanded in terms of these discrete operators
\begin{equation}
\pmb{\Phi} (x,t)= \sum_{n=-\infty}^\infty \pmb{\Phi}_s^k (n,t) s^k_n (x) +
\sum_{n=-\infty}^\infty\sum_{l=k}^\infty
\pmb{\Phi}_w^l (n,t) w^l_n (x), 
\label{a:40}
\end{equation}
with a similar expression for $\pmb{\Pi} (x,t)$.
These expansions decompose the fields into well-defined operators
associated with different resolutions.

We can construct resolution $m$ truncations of these fields by
discarding degrees of freedom associated with scales smaller than
$1/2^m$:
\begin{equation}
\pmb{\Phi}^{m} (x,t) = \sum_{n=-\infty}^{\infty} \pmb{\Phi}_s^{m} (n,t)
s_n^m (x) =
\sum_{n=-\infty}^{\infty} \pmb{\Phi}_s^{k} (n,t)
s_n^k (x) +
\sum_{n=-\infty}^{\infty}\sum_{l=k}^{{m-1}} \pmb{\Phi}_w^{l} (n,t)
w_n^l (x),
\label{a:42}
\end{equation}
again with an analogous expansion for $\pmb{\Pi}^{m} (x,t)$. 
Truncations on $n$ give volume cutoffs.  The advantage of having a
basis is that the eliminated degrees of freedom can be
systematically restored.

It is possible to replace the discrete Hermitian canonical fields by
discrete creation and annihilation operators
\begin{equation}
\mathbf{a}_s^k({n},t) : = {1 \over \sqrt{2} } 
(\sqrt{\gamma} \pmb{\Phi}_s^k ({n},t) +i {1 \over \sqrt{\gamma}} 
\pmb{\Pi}_s^k ({n},t)), 
\end{equation}
\begin{equation}
\mathbf{a}_w^k({n}, t) : ={1 \over \sqrt{2} }((\sqrt{\gamma}  
\pmb{\Phi}_w^k ({n},t)+i {1 \over \sqrt{\gamma}} \pmb{\Pi}_w^k ({n},t))
\label{a:45}
\end{equation}
where $\gamma$ is chosen so 
$\mathbf{a}_s^k({n},t)$ and 
$\mathbf{a}_w^k({n}, t)$ annihilate the vacuum.


One of the difficulties with field theories is that products of local
fields at the same point are ill-defined.  On the other hand products
of smeared fields are well defined, but because of the smearing they
lose their local character. Replacing the fields in the Hamiltonain
or Poincar\'e generators by expansions of the form (\ref{a:40}) 
means that the products of the individual operators are well
defined; but because we are writing the exact ill-behaved Hamiltonian
as a sum of well-defined operators means that these sums will not
converge.  On the other hand, if the fields are replaced by finite
volume - finite resolution truncations of the type
(\ref{a:42}),
then the truncated theory is well defined.

It is instructive to exhibit the structure of 
the resolution $1/2^k$ Hamiltonian for a 1+1 dimensional $:\phi^4(x):$ 
interaction. It has the form
\[
H^k=  
{1 \over 2}\sum : \left ( \pmb{{\Pi}}^k_s(n,0)^2 +
D^k_{mn} \pmb{\Phi}^k_s (n,0) \pmb{\Phi}^k_s (m,0)
+   \mu^2  \pmb{\Phi}^k_s(n,0)^2 + \right . 
\]
\begin{equation}
\left . \lambda  \Gamma^k_{n_1 n_2 n_3 n_4} 
\pmb{\Phi}^k_s (n_1,0)
\pmb{\Phi}^k_s (n_2,0)
\pmb{\Phi}^k_s (n_3,0)
\pmb{\Phi}^k_s (n_4,0) \right ):
\label{a:50}
\end{equation}
where the numerical coefficients $D^k_{mn}$ and $\Gamma^k_{n_1 n_2 n_3 n_4}$
are overlap integrals of derivatives of scaling functions or
products of scaling functions: 
\begin{equation}
D^k_{mn}= \int dx {\partial \over \partial x} s^k_m (x)
{\partial \over \partial x} s^k_n (x), \qquad
\Gamma^{k}_{n_1 n_2 n_3 n_4} := \int s_{n_1}^k(x)
s_{n_2}^k (x)s_{n_3}^k (x) s_{n_4}^k(x)  dx .
\label{a:52}
\end{equation}

The advantage of using basis functions that are related to fixed points of 
a renormalization group equation is that the Hamiltonians with different 
resolutions have the same form. The only difference is that the numerical 
coefficients (\ref{a:52}) for different scale truncated 
Hamiltonians differ by powers of $2$.  For the coefficients 
(\ref{a:52}) these scaling identities are 
\begin{equation}
D^k_{mn}= 2^k D^0_{mn} , \qquad 
\Gamma^{k}_{n_1 \cdots n_m} = 2^{k(m-2)/2} \Gamma^{0}_{n_1 \cdots n_m}
\label{a:54}
\end{equation}
so it is only necessary to know these quantities on one scale. 

The renormalization group equations imply that scale 1 quantities 
satisfy finite systems of algebraic equations 
\begin{equation}
\Gamma^0_{0 n_2 n_3 } =
\sqrt{2}  
\sum h_{l_1} h_{l_2} h_{l_3} \Gamma^0_{0 2n_2+l_2-l_1, 2n_3+l_3-l_1 } \qquad
\sum_{n_3} \Gamma^0_{0 n_2 n_3 } = \delta_{n_2 0}
\label{a:56}
\end{equation}
\begin{equation}
D^0_{0, n_1} = \sum  4 h_{l_1}h_{l_2}  D^0_{0,l_2n_2 +l_2-l_1}
\label{a:57}
\end{equation}
which can be solved for all of the coefficients that appear in the 
truncated Hamiltonians at any scale\cite{ref3}.

An important identity is the relation between the discrete 
field operators on adjacent scales, given by 
\begin{equation}
\pmb{\Phi}^{k+1}_s (n,0) =
\sum_m (h_{n-2m} \pmb{\Phi}^{k}_s (n,0) +
g_{n-2m} \pmb{\Phi}^{k}_w (n,0) ).
\label{a:57}
\end{equation}

Using (\ref{a:45}) and (\ref{a:57}) in the scale $k+1$ Hamiltonian
(\ref{a:50}) gives
\begin{equation}
H^{k+1}_s (\mathbf{a}^{k+1}_s,\mathbf{a}^{k+1}_s{}^{\dagger})  = 
H^{k}_s (\mathbf{a}^{k}_s,\mathbf{a}^{k}_s{}^{\dagger}) + H^{k}_w 
(\mathbf{a}^{k}_w,\mathbf{a}^{k}_w{}^{\dagger})
+ H^{k}_{sw}(\mathbf{a}^{k}_s,\mathbf{a}^{k}_s{}^{\dagger},
\mathbf{a}^{k}_w,\mathbf{a}^{k}_w{}^{\dagger}) 
\label{a:58}
\end{equation}
where $H^{k+1}_s (\mathbf{a}^{k+1}_s,\mathbf{a}^{k+1}_s{}^{\dagger})$ 
and $H^{k}_s (\mathbf{a}^{k}_s,\mathbf{a}^{k}_s{}^{\dagger})$ 
both have the form
(\ref{a:50}).   
$H^{k}_w 
(\mathbf{a}^{k}_w,\mathbf{a}^{k}_w{}^{\dagger})$ represents fine-scale 
degrees of freedom that are not coupled to the coarse-scale Hamiltonian, 
while $ 
H^{k}_{sw}(\mathbf{a}^{k}_s,\mathbf{a}^{k}_s{}^{\dagger},
\mathbf{a}^{k}_w,\mathbf{a}^{k}_w{}^{\dagger})$ includes the operators that 
couple the two scales. 

In these expression the coarse and fine-scale Hamiltonians have the
same form.  The coarse scale Hamiltonain is fixed by adjusting the
bare masses and coupling parameters to agree with some coarse scale
observables.  If one is only interested in coarse-scale observables it
is possible to construct a unitary operator that decouples the coarse
and fine scale degrees of freedom.  This gives a coarse-scale
Hamiltonain that involves only explicit coarse-scale degrees of
freedom, but has coefficients that include the effects of the
eliminated fine scale degrees of freedom.  The new coarse-scale
Hamiltonian involves the same parameters (bare masses and coupling
constants) as the original coarse-scale Hamiltonain.  These have to be
re-adjusted in order to keep the value of the coarse scale observables
unchanged.  This can in principle be repeated; at each stage the bare
observables need to be adjusted as the effects of additional
fine-scale degrees of freedom are included.  By absorbing some of the
scaling behavior in the mass and coupling constants, one gets
renormalized parameters.  One gets non-trivial theories if this
process leads to finite limits of the renormalized quantities in the
limit that the effects of arbitrarily fine scale observables are
included.
 
The elimination of the small-scale degrees of freedom can be attempted 
using a number of methods, such similarity renormalization group methods
\begin{equation}
{dH(\lambda) \over d \lambda} = 
[H(\lambda),[H(\lambda), G] ] 
\label{a:59}
\end{equation}
where $G$ is a generator that is chosen to evolve the Hamiltonian to
a form that decouples the low and high-resolution degrees of freedom.
An advantage in any of these methods is that all of the operators
have the structure of known constants with known scaling properties
and discrete creation and annihilation operators, so everything is
easily computable.

Another important feature of the wavelet truncation of fields 
takes advantage of the partition of unity property.  
Noether's theorem gives formal expression for the Poincar\'e generators 
in terms of local densities at a fixed time, satisfying   
\begin{equation}
[ O^a(x) , O^b(y) ] = i \delta (x-y ) f^{abc}O^c (y) 
\label{a:60}
\end{equation}
where $f^{abc}$ are the structure constants of the Poincar\'e Lie algebra. 
Using the partition of unity in the forms 
\begin{equation}
1= (2^{-k/2} \sum_n s^k_{n} (x) )
(2^{-k/2} \sum_m s^k_{m} (y)) \qquad 
1= (2^{-k/2} \sum_n s^k_{n} (x) )
\label{a:62}
\end{equation}
on the left and right respectively,  gives the exact identity 
\begin{equation} 
[\sum_n O^{ak}_n, \sum_m O^{bk}_m] = 
i f^{abc} \sum_l O^{ck}_l
\label{a:64}
\end{equation}
where 
\begin{equation}
O^{ak}_n := 2^{k/2} \int O^a (x) s^k_{n}(x) dx .
\label{a:65}
\end{equation}

On the other hand, since $O^a (x)$ involves product of fields,
if each field in the product is replaced by the finite resolution 
approximations,
\begin{equation}
\pmb{\Phi} (\mathbf{x},t) \to \pmb{\Phi}^{k} (\mathbf{x},t) 
\qquad \pmb{\Pi} (\mathbf{x},t) \to \pmb{\Pi}^{k} (\mathbf{x},t), 
\label{a:66}
\end{equation} 
then $O^{ak}_n$ will be replaced by an approximation that will not
satisfy equation (\ref{a:64}). Obviously the correct
commutation relations must be recovered in the infinite resolution
limit.  These violations of the symmetry can be studied by looking at the 
scale of the terms that violate the commutation relations.


\begin{thebibliography}{3}
%
%

\bibitem{ref1} I. Daubechies, (1988) 
Comm. Pure Appl. Math. {\bf 41}, 909.

\bibitem{ref2} I. Daubechies, 
{\it Ten Lectures on Wavelets},
SIAM, Philadelphia, 1992.

\bibitem{ref3} F. Bulut and W. N. Polyzou, (2013) 
Phys. Rev. {\bf D}87, 116011.

\bibitem{ref4} Christoph Best, Andreas Schaefer,
arXiv: hep-lat/9402012, 1994.

\bibitem{ref5} Christoph Best, (2000)
Nucl. Phys. Proc. Suppl. {\bf 83},848.

\bibitem{ref6} Ahmed E. Ismail, Gregory C. Rutledge, 
and George Stephanopoulos, (2003)  
J. Chem. Phys. {\bf 118},4414.

\bibitem{ref7} Ahmed E. Ismail, Gregory C. Rutledge, 
and George Stephanopoulos, (2003)  
J. Chem. Phys. {\bf 118},4424.

\bibitem{ref8} I.G. Halliday, P. Suranyi, (1995)
Nuc. Phys. B436,414.

\bibitem{ref9} P. Federbush, (1995)
Prog. Theor. Phys. {\bf 94},1135.

\bibitem{ref10} Guy Battle, {\it Wavelets and Renormalization}, 
Series in Approximations and Decompositions, Volume 10, 
World Scientific, 1999.

\bibitem{ref11} Mikhail V. Altaisky, (2007)
SIGMA {\bf 3},105.

\bibitem{ref12} S. Albeverio, Mikhail V. Altaisky, 
arXiv:0901.2806v2,2009.

\bibitem{ref13} Mikhail V. Altaisky, (2010)
Phys. Rev. D {\bf 81},125003.

\end{thebibliography}
\end{document}